# Regional Deep Atrophy: a Self-Supervised Learning Method to Automatically Identify Regions Associated With Alzheimer's Disease Progression From Longitudinal MRI

Mengjin Dong [a, #], Long Xie [a, b], Sandhitsu R. Das [a, c, d], Jiancong Wang [a], Laura E.M. Wisse [a, b, e], Robin deFlores [c, d, f], David A. Wolk [c, d], Paul A. Yushkevich [a, b, #], for the Alzheimer's Disease Neuroimaging Initiative [*] a Penn Image Computing and Science Laboratory (PICSL), Department of Radiology, University of Pennsylvania, Philadelphia, PA, United States
b Department of Radiology, University of Pennsylvania, Philadelphia, Pennsylvania, United States
c Department of Neurology, University of Pennsylvania, Philadelphia, Pennsylvania, United States
d Penn Memory Center, University of Pennsylvania, Philadelphia, Pennsylvania, United States
e Department of Diagnostic Radiology, Lund University, Lund, Sweden
f Institut National de la Santé et de la Recherche Médicale (INSERM), Caen, France

## Abstract

Longitudinal assessment of brain atrophy, particularly in the hippocampus, is a well-studied biomarker for neurodegenerative diseases, such as Alzheimer's disease (AD). In clinical trials, estimation of brain progressive rates can be applied to track therapeutic efficacy of disease modifying treatments. However, most state-of-the-art measurements calculate changes directly by segmentation and/or deformable registration of MRI images, and may misreport head motion or MRI artifacts as neurodegeneration, impacting their accuracy. In our previous study, we developed a deep learning method DeepAtrophy that uses a convolutional neural network to quantify differences between longitudinal MRI scan pairs that are associated with time. DeepAtrophy has high accuracy in inferring temporal information from longitudinal MRI scans, such as temporal order or relative inter-scan interval. DeepAtrophy also provides an overall atrophy score that was shown to perform well as a potential biomarker of disease progression and treatment efficacy. However, DeepAtrophy is not interpretable, and it is unclear what changes in the MRI contribute to progression measurements. In this paper, we propose Regional Deep Atrophy (RDA), which combines the temporal inference approach from DeepAtrophy with a deformable registration neural network and attention mechanism that highlights regions in the MRI image where longitudinal changes are contributing to temporal inference. RDA has similar prediction accuracy as DeepAtrophy, but its additional interpretability makes it more acceptable for use in clinical settings, and may lead to more sensitive biomarkers for disease monitoring in clinical trials of early AD.

---

[*] Data used in preparation of this article were obtained from the Alzheimer's Disease Neuroimaging Initiative (ADNI) database (adni.loni.usc.edu). As such, the investigators within the ADNI contributed to the design and implementation of ADNI and/or provided data but did not participate in analysis or writing of this report. A complete listing of ADNI investigators can be found at: http://adni.loni.usc.edu/wp-content/uploads/how_to_apply/ADNI_Acknowledgement_List.pdf

# Introduction

Long before the emergence of symptoms, individuals with Alzheimer's disease (AD) tend to have accelerated neuronal loss compared to unaffected individuals of similar age. This accelerated neuronal loss is shown to start in medial temporal lobe (MTL) region, and then gradually spreads out to the whole brain (Braak & Braak, 1995; Giannakopoulos et al., 2003; Gunten et al., 2006). Symptoms such as cognitive impairment or memory loss will appear in later stages of AD. Recently approved treatments Lecanemab (Swanson et al., 2021) and Aducanumab (Tolar et al., 2020), as well as many experimental treatments in clinical trials, target modifying the disease process at an early stage, before extensive neuronal damage has occurred, and symptoms of dementia have developed. The effectiveness of these treatments can be estimated by observing difference of disease progression rate from MRI images between the treatment group and the age-matched non-AD group. Specifically, longitudinal measurement of change in hippocampal volume, as a part of the MTL, is a well-studied biomarker to track disease progression. Compared to cognitive tests, biomarkers derived from longitudinal structural MRI are more sensitive to disease progression in early AD, as demonstrated in prior studies that compared longitudinal MRI and cognitive measures between groups with different levels of AD severity and cognitively unimpaired controls (Caroli & Frisoni, 2010; Hua et al., 2009).

Multiple methods have been developed to measure longitudinal volume changes of brain structures from the same subject. A natural way to measure longitudinal volume change is to quantify volumes in each scan from the same subject independently and compare the volume difference. However, this method turns out to be susceptible to small errors in each measurement and generate change measurements with large variance (Das et al., 2012). Recent work has favored an alternative approach, where pairs of longitudinal MRI scans are analyzed jointly and change in volume is quantified directly, typically via deformable registration. As is shown in a recent challenge presented in (Cash et al. 2015), there were two dominant approaches to longitudinal volume change analysis based on image registration. One class of methods is *deformation-based morphometry (DBM)*, also known as *tensor-based morphometry (TBM)* (Das et al., 2012; Hua et al., 2008; Reuter et al., 2012; Xie, Wang, et al., 2019). In DBM, deformable registration is applied to a pair of longitudinal MRI scans, and a deformation field depicting the spatial transformation from the *baseline image* (term used to refer to the first scan in a longitudinal study) to the *follow-up image* (term used to refer to later scans in a longitudinal study) at each voxel is obtained. After calculating the segmentation mask and volume of the hippocampus (or another structure of interest) in the baseline image, the deformation field is applied to this segmentation to obtain the follow-up volume of the hippocampus. Atrophy rate of the hippocampus is then calculated as the normalized ratio of the volumetric difference between baseline segmentation and follow-up segmentation and the time interval between scans. The other class of approaches is *Boundary Shift Integral (BSI)*

(Gunter et al., 2003; Leung et al., 2009; Prados et al., 2014). Rather than calculating a voxel-wise deformation field in DBM, BSI focuses on intensity difference along the boundaries of specific regions of interest (ROIs), and as the name of this algorithm suggests, the hippocampus boundary shift, to calculate volumetric change. However, both classes of registration-based techniques implicitly assume that any deformation detected between a pair of longitudinal scans is attributable to biological structural changes, whereas in practice, measured deformations may also be caused by factors such as motion, imaging artifacts, head position difference and scanner difference, which can confound the results of atrophy estimation.

Recent image registration methods based on deep learning (Dalca et al., 2019; Fu et al., 2020; Mok & Chung, 2020) show results comparable to numerical registration methods, while significantly reducing inference time, and are hence an attractive alternative for quantification of longitudinal changes in MRI. However, deep learning-based registration models commonly have the same optimization objective as conventional registration methods, i.e. to minimize image dissimilarity while regularizing the transformation field. Therefore, simply applying deep learning-based registration models in the context of DBM is unlikely to help differentiate biological structural changes from unrelated factors.

Most DL-based analyses of longitudinal or cross-sectional MRI in AD are more focused on automatic diagnosis (Hao et al., 2020; Lee et al., 2019), prediction of future disease progression (Ghazi et al., 2019; Gibson et al., 2018; Q. Li et al., 2019), or cognitive score prediction (Lu et al., 2021; Nguyen et al., 2020). In one of the recent longitudinal MRI challenges, TADPOLE challenge (Răzvan V. Marinescu et al., 2019), three objectives were defined for longitudinal image prediction: AD diagnosis, cognitive score, and ventricle volume for each subject. In this challenge, most highly ranked teams are based on XGBoost (Razvan V Marinescu et al., 2020) and statistical regression models (Aksman et al., 2019). However, these objectives are distinct from the goal of our paper, which is to develop a biomarker based on longitudinal MRI that is as sensitive as possible to differences in disease progression between individuals; a goal of high relevance to clinical trials of disease-modifying treatments in AD, as well as to clinical application of recently approved disease-modifying treatments. To our knowledge, only a few papers use deep learning and self-supervised learning methods to extract features that can explicitly represent structural progression of AD from MRI images. In Ouyang et al. (2022), the authors used self-supervised learning methods to constrain over the embedding vector space of longitudinal image pairs, and generate embedding vectors representative of progression related to aging. In Ren et al. (2022), a longitudinally-consistent spatiotemporal representation learning framework was proposed for longitudinal segmentation and spatial temporal consistency evaluation. However, both studies were based on downsampled whole brain images, as opposed to our approach, which explicitly focuses on the MTL, a region of the brain where earliest AD-related neurodegeneration occurs, and which attempts to isolate changes in longitudinal MRI that are truly driven by neurodegeneration, rather than unrelated factors.

Our previous work DeepAtrophy (Dong et al., 2021) used deep learning methods to infer a relative progression rate from a pair of longitudinal MRI scans, which is compared to the progression rate of normal aging group. Previous studies show that disease groups tend to exhibit more progressive changes (atrophy) than healthy control groups over the same period. However, DeepAtrophy's prediction of relative interscan intervals assumes a uniform rate of progression for all subjects, regardless of their disease status. As a result, if an individual experiences accelerated brain atrophy, such as in the case of an AD patient, the predicted interscan interval is expected to be longer than actual interscan intervals. The inference of DeepAtrophy is conducted by two auxiliary objectives used when training deep learning models: for a pair of images input to the model in an arbitrary order, determine the correct scan temporal order (STO); and for two pairs of images with the scan time interval of one pair strictly inside the other pair, also arbitrarily input to the model, determine which pair has a longer relative interscan interval (RISI). In other words, when training the DeepAtrophy model, we input sets of scans of the same individual in an arbitrary order and ask the model to infer information about time order based on image content. Since neurodegeneration is associated with the passage of time, while other factors, such as imaging artifacts, are more likely to be independent of time, DeepAtrophy is likely implicitly detecting neurodegeneration by learning to correctly infer temporal information from longitudinal MRI scans. The results of DeepAtrophy show good accuracy in inferring scan temporal order and relative interscan interval on held-out test data. Significantly, DeepAtrophy detects an accelerated progression rate in a preclinical AD group (i.e., cognitively unimpaired individuals with evidence of AD pathology based on a positive β-amyloid positron emission tomography (PET) scan, A+ CU) compared to a control group of β-amyloid PET negative cognitively unimpaired individuals (A-CU), using scans within only two years of baseline for each individual, demonstrating a potential to be a surrogate biomarker to estimate disease progression in future clinical trials. However, DeepAtrophy only generates a scalar relative progression score for each pair of longitudinal MRI scans, which cannot be directly attributed to atrophy in any specific anatomical structure or region. This is a significant limitation, since in clinical trials and other clinical applications, model interpretability and transparency are important considerations.

In this paper, we propose a new method called Regional Deep Atrophy (RDA). Similar to DeepAtrophy, RDA takes as input pairs of longitudinal MRI scans from the same subject and seeks to distinguish changes that are progressive (i.e., truly related to neurodegeneration) from changes that are due to spurious factors, such as different patterns of head motion or MRI artifact in the two scans. Like DeepAtrophy, RDA is trained using loss functions based on temporal information (STO, RISI) to help the model distinguish between progressive and spurious changes. However, whereas DeepAtrophy operates as a "black box" that outputs a single measure of progressive change, and hence lacks interpretability, RDA is designed to mimic a conventional DBM approach in that it computes a deformation field between pairs of scans and bases its longitudinal change measure on applying this deformation field to regions of interest (ROIs). This deformation field can be generated by any image registration model and is fixed in RDA training. However, rather than focusing on a predetermined structure like the hippocampus, RDA employs an attention mechanism to identify ROIs where deformation is linked to progressive change. This attention model uses a segmentation-like network (a 3D U-net) to generate ROIs associated with shrinkage (as would be expected for grey matter structures) and expansion (as would be expected for areas filled with cerebrospinal fluid). The overall measure of change is computed by quantifying the change in volume induced on the shrinkage and expansion ROIs by the deformation field and combining these quantities with opposite signs. The underlying expectation is that areas of the image most affected by spurious factors (such as head motion) would be excluded from the shrinkage/expansion ROIs proposed by the attention network, increasing the overall sensitivity of the change measures to neurodegeneration.

The primary contributions of this paper are to propose a technique that combines the ability of DeepAtrophy to disentangle progressive vs. spurious changes in longitudinal MRI scan pairs with the interpretability of DBM; and to evaluate this technique using a large longitudinal MRI dataset from ADNI (Mueller et al., 2005). The registration branch of RDA consists of the deep learning-based deformable image registration model VoxelMorph (Dalca et al., 2019). Since, to our knowledge, VoxelMorph has not been previously evaluated in the context of brain MRI longitudinal change detection, a secondary contribution of this paper (Section 2.2, Section 3.1-3.4) is to evaluate VoxelMorph in this context, comparing it to a DBM pipeline based on a conventional registration algorithm (Das et al., 2012).

## Material and methods

The overall structure of RDA network is shown in Figure 1. RDA consists of two branches, the deformable registration branch (VoxelMorph), which is pre-trained and used with fixed weights during RDA training; and the attention branch, which uses a U-net architecture. Pairs of longitudinal MRI scans of the same subject (more precisely, ROIs cropped around the left or right MTL and globally aligned) are input to both branches. The temporal order of the inputs is arbitrary, i.e., 50% of the time the first input is the earlier scan and 50% of the time the second input is the earlier scan. VoxelMorph uses these inputs to generate a smooth deformation field that matches corresponding locations between the two images, and the attention branch generates ROIs corresponding to putative regions of expansion (shown in purple) and shrinkage (shown in green). Applying the deformation field to these ROIs yields summary measures of expansion and shrinkage, which can be combined with opposite signs to get an overall change measure. The STO loss is used to evaluate how well summary measures predict *scan temporal order*, i.e., classify scan pairs input in correct temporal order vs. scan pairs input in reverse temporal order. The RISI loss evaluates how well this measure can predict *relative interscan intervals* between two pairs of scans from the same subject, i.e., distinguish two scans with a shorter interval from two scans with a longer interval. When computing the RISI loss, the two pairs of scans are passed as inputs to two copies of the RDA network (with shared weights), and their outputs are used in the RISI loss computation (Figure 1).

The structure of this section is as follows: Section 2.1 introduces the dataset and pre-processing procedures used in our experiments; Section 2.2 describes VoxelMorph model training and evaluation; Section 2.3 details the design of the RDA network and the STO and RISI losses; Section 2.4 introduces the evaluation metrics used in this study and the algorithms used for comparison; lastly, some hyperparameters and implementation details are specified in Section 2.5.

## Participants and pre-processing

Data used in this study were obtained from the Alzheimer's Disease Neuroimaging Initiative (ADNI, adni.loni.usc.edu). The ADNI was launched in 2003 as a public-private partnership, led by Principal Investigator Michael W. Weiner, MD. The primary goal of ADNI has been to test whether serial magnetic resonance imaging (MRI), positron emission tomography (PET), other biological markers, and clinical and neuropsychological assessment can be combined to measure the progression of mild cognitive impairment and early Alzheimer's disease. For up-to-date information, see www.adni-info.org.

502 Participants from ADNI2 and ADNI GO were included in this study. All participants have 2-6 T1-weighted MRI scans, spanning from 0.25 to 6 years. In total, 4927 pairs of same-subject MRI scans are included in this dataset. A summary standardized uptake value ratio (SUVR) derived from associated Florbetapir PET images collected within 0.5 years of baseline MRI scan were used to determine β-amyloid (A) status (threshold of 1.11; Landau et al., 2012). Participants were grouped into four cohorts corresponding to progressive stages along the AD continuum: A- cognitively unimpaired older adults (A-CU), A+ cognitively unimpaired older adults (A+CU, also referred to as the *preclinical AD* group), and A+ early and late Mild Cognitive Impairment (A+ eMCI and A+ lMCI).

Participants were divided into training (n = 155), validation (n = 21), and test (n = 326) groups. 72 subjects with only two MRI scans were assigned to the test group since in the training stage at least three scans are required. All 85 preclinical AD subjects (A- CU) were also assigned to the test group to keep a large enough sample size for the group difference analysis. The rest of the subjects (347) were split into training, validation, and test groups by a proportion of 45%, 5%, and 50% (numbers were rounded to the nearest integer). Demographics such as age, sex, and Mini-Mental State Examination (MMSE) scores of all four groups in the training and test set are included in Table 1.

The same preprocessing steps as in DeepAtrophy (Dong et al., 2021) were applied to our dataset. First, the ADNI T1-MRI scans were up-sampled by a factor of two in the coronal plane using non-local mean super-resolution technique (Coupé et al., 2013; Manjón et al., 2010). Next, the left and right MTL were segmented using ASHS-T1 segmentation software (Xie, Wisse, et al., 2019) and image regions around the left and right MTL were cropped with a margin of ~10 voxels. These left and right MTL images were treated separately in subsequent analysis, and the MTL segmentations were not used in the RDA model other than for cropping the MTL region. However, ASHS-T1 MTL segmentations were used by the DBM techniques for comparison. Next, rigid alignment was performed separately for the left and right MTL region in each pair of longitudinal scans. The rigid transformation was decomposed into two equal halves (i.e., $R = R_{half} \circ R_{half}$), and these halves were applied to the fixed and moving images, so that the amount of aliasing after applying the rigid transformation would be equivalent for the fixed and moving images (Yushkevich et al.,2009). Each pair of scans was registered twice, with the earlier scan as the fixed image and the later scan as the moving image, and vice versa. Overall, 19,708 pairs of 3D images were analyzed (4927 scan pairs * 2 sides * 2 registration directions). Since manually checking this number of image pairs for registration errors is impractical, we computed the Structural SIMilarity (SSIM) score (Wang et al., 2004) between registered scan pairs and excluded all scan pairs with SSIM < 0.6. Visual examination of scan pairs shows that the SSIM of most successfully registered 3D scan pairs are between 0.7 and 0.9, and an SSIM < 0.6 of an image pair would mean almost always strong misalignment between two images. A total of 1414 scan pairs (7.2%) was rejected through this procedure.

Table 1. Characteristics of the selected ADNI2/GO participants whose T1 MRI scans were used for the Regional Deep Atrophy (RDA) and comparison experiments for this paper. Numbers in parentheses are standard deviations. All subjects in the training and test set had 2-6 scans between 0.25 and 6 years from the baseline. Abbreviations: n = number of subjects; A+/A-: β-amyloid positive/negative; CU = cognitively unimpaired adults; eMCI = early mild cognitive impairment; lMCI = late mild cognitive impair; Edu = years of education; MMSE = mini-mental state examination.

|  | Training Set (n = 155) | | | | Test Set (n = 326) | | | |
| --- | --- | --- | --- | --- | --- | --- | --- | --- |
|  | A- CU (n = 64) | A+ CU (n = 0) | A+ eMCI (n = 49) | A+ lMCI (n = 42) | A- CU (n = 103) | A+ CU (n = 85) | A+ eMCI (n = 79) | A+ lMCI (n = 59) |
| Age | 71.5 (5.7) | - | 73.3 (7.4) | 71.9 (7.2) | 72.4 (6.3) | 75.2 (5.8)** | 74.0 (6.9) | 72.5 (6.5) |
| Sex | 27F 37M | - | 16F 33M | 21F 21M | 56F 47M | 58F 27M | 39F 40M | 28F 31M |
| Edu | 17.4 (2.3) | - | 16.2 (3.0)* | 16.8 (2.7) | 16.8 (2.4) | 16.0 (2.7) | 15.3 (2.9)*** | 16.4 (2.7) |
| MMSE | 29.1 (1.2) | - | 28.0 (1.5)**** | 26.7 (1.8)**** | 29.1 (1.2) | 29.0 (1.1) | 28.1 (1.6)**** | 27.3 (1.9)**** |

*Notes*: All statistics in the train and test set are in comparison to the corresponding A- CU group. *SD* is reported in parenthesis. Independent two-sample *t*-test (continuous variables with normal distribution, for age and education), Mann–Whitney *U* test (continuous variable with nonnormal distribution, for MMSE) and contingency $\chi^2$ test (sex) were performed.
Abbreviations: *, p < 0.05; **, p < 0.01; ***, p < 0.001; ****, p < 0.0001.

### Image Registration between Longitudinal Pairs using VoxelMorph

In our model, we first train a registration model to obtain local deformation fields between longitudinal image pairs. Deformation fields between longitudinal images capture changes in anatomy in interpretable terms; for instance, the determinant of the Jacobian matrix of the deformation field provides a local measure of shrinkage or expansion of tissue over time. We adopted VoxelMorph (Dalca et al., 2019) in our model because (1) it is a deep learning-based image registration model that can generate a deformation field with only one pass, while retaining similar accuracy as the orders of magnitude more computationally expensive optimization-based registration methods (Avants et al., 2007; Das et al., 2012) used in conventional DBM; (2) VoxelMorph can be more easily integrated as a component in our deep learning pipeline compared to optimization-based registration methods.

VoxelMorph combines the U-Net (Ronneberger et al., 2015) architecture with a scaling and squaring layer (Arsigny et al., 2006) to transform input image pairs (consisting of a fixed image and a globally aligned moving image) into a deformation field of the same dimensions as the input images. During training, the U-Net weights are updated to minimize a weighted sum of two loss terms analogous to the objective functions used by conventional registration methods (Avants et al., 2007; Beg et al., 2005; Klein et al., 2010) – a data similarity term and a regularization term. The data similarity term helps the model to reach maximum similarity between the warped moving image and the fixed image, by minimizing a local dissimilarity metric Normalized Cross Correlation (NCC); and the regularization term minimizes the L2 norm of the generated deformation field. Together, the regularization term and the scaling and squaring layers in VoxelMorph help the model to generate a smooth and invertible deformation field, which is desirable in many medical image registration tasks, since a one-to-one mapping between the moving and fixed images often reflects a physically plausible tissue change in biological structures.

VoxelMorph model has been successfully applied or adapted to multiple image registration tasks, such as brain MR images (J. Chen et al., 2022; Fan et al., 2019a), and abdomen CT images (H. Tang et al., 2021), achieving matching accuracy and deformation field regularity comparable to conventional methods but with a much faster registration time. However, to our knowledge, there has been little work to evaluate VoxelMorph in the context of longitudinal registration in DBM, where the changes are more localized and subtle than in between-subject registration problems. Hoopes et al., (2021) evaluated VoxelMorph registration accuracy in longitudinal brain MRI from ADNI, but did not derive or evaluate atrophy measures from it. Therefore, in addition to integrating VoxelMorph as a component in our RDA model (in Sections 2.3 and 2.4), we implemented a DBM pipeline based on VoxelMorph and compared its performance to the conventional DBM pipeline Automatic Longitudinal Hippocampal Atrophy software/package (ALOHA) (Das et al., 2012), which remains a state-of-the-art hippocampus atrophy estimation model to date.

We trained VoxelMorph using the train/evaluate/test split described in Section 2.1. The results from a recent paper by (Hoopes et al., 2021) on automatic selection of hyper-parameters showed that the weights of the data term and regularization term reached an optimal for intra-subject registration tasks on ADNI dataset when set as 0.85 and 0.15, respectively. Data augmentation included random flipping, rotation, normalization, and cropping around the MTL segmentation. To integrate VoxelMorph into a DBM pipeline, we simply replaced the conventional deformable registration step in the ALOHA pipeline with VoxelMorph. Specifically, after applying the MTL segmentation, cropping, and rigid alignment steps described in Section 2.1, pairs of same-subject scans from the test set were input to VoxelMorph and the deformation field was applied to the ASHS-T1 hippocampus segmentation in the fixed image to estimate the hippocampus volume in the moving image. The annualized rate of change in hippocampus volume is then derived by taking the difference between fixed and moving hippocampus volumes and dividing it by the fixed hippocampus volume and interscan interval. Estimation of annualized rate of change when multiple pairs of scans are available for a subject is discussed in Section 2.5.

Figure 1. Overall architecture of Regional Deep Atrophy (RDA) pipeline. Two pairs of images of the same subject (pairs $(I_a, I_b)$ and $(I_a, I_c)$ in this example) are input to two copies of the RDA network in an arbitrary order. For each copy, the input image pair is passed in the same order to two sub-networks, a pretrained VoxelMorph network and a U-net-like Attention Network. A change in volume is calculated from the deformation field generated by VoxelMorph and the shrinkage/expansion attention maps generated by the Attention Network (with green area representing shrinkage region and purple area representing expansion region). Scan Temporal Order (STO) loss was calculated based on whether the sign of total volume change is consistent with the ordering of the input image pair, and Relative Interscan Interval (RISI) loss was calculated based on whether a larger total volume change is consistent with a longer interscan interval in two image pairs. In the test stage, only a longitudinal image pair is required and the total volume change is calculated as the final output for this pair, without STO or RISI loss calculation.

Training stage:

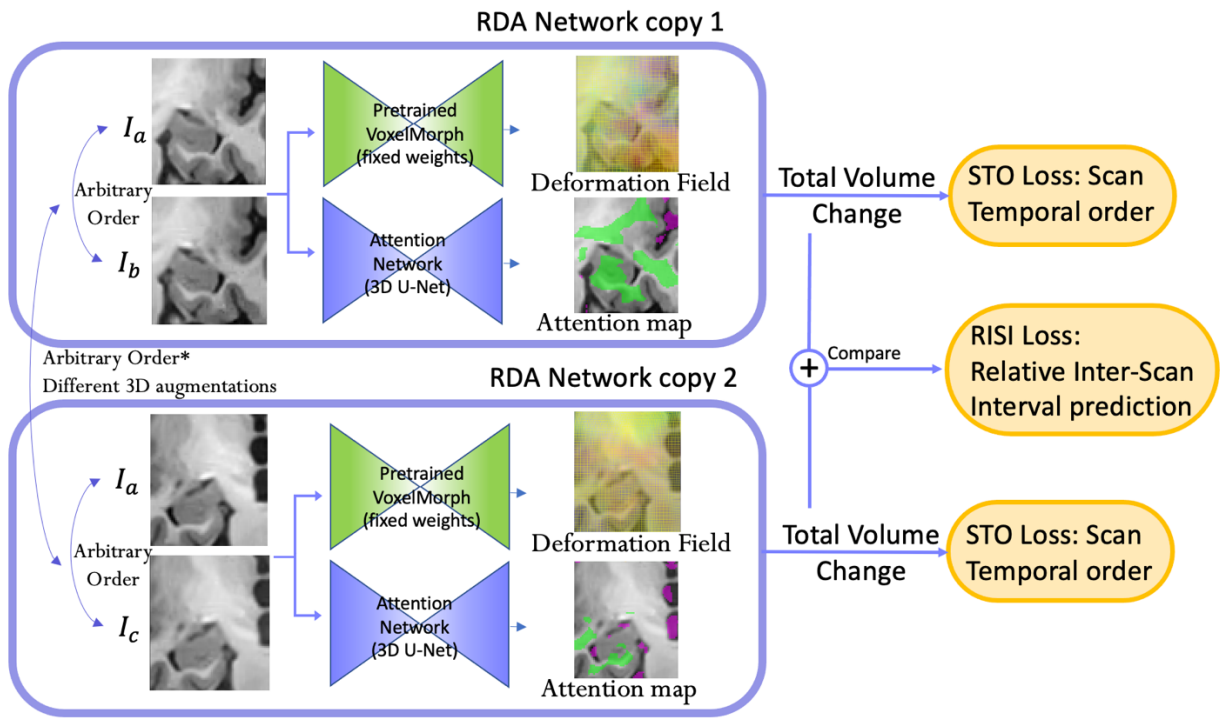

Test stage:

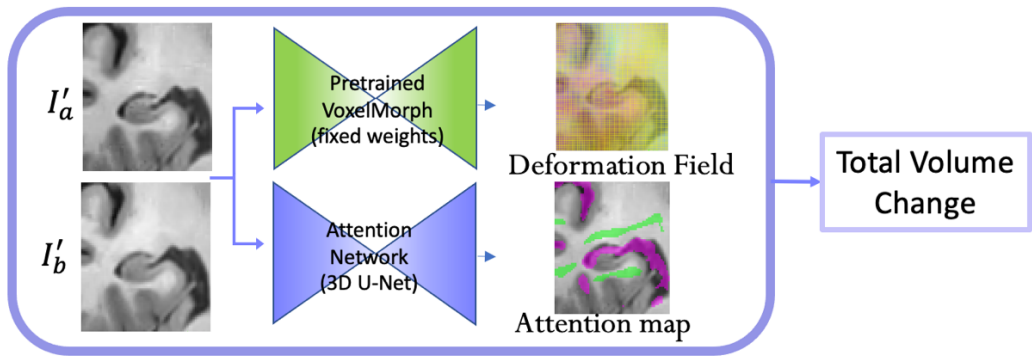

## RDA Model architecture – Attention model and time-related losses

Similar to DeepAtrophy (Dong et al., 2021), we make a strong simplifying assumption that in older adults, the change in the volume of brain structures is monotonic with respect to time, i.e., the volume of grey and white matter structures, such as the hippocampus, decreases over time and the volume of fluid structures, such as the lateral ventricles, increases over time. Moreover, we assume that the amount of change in brain structure measured from two MRI images of the same subject $j$ at different time points $I^{j,t_1}$ and $I^{j,t_2}$ should be approximately proportional to the inter-scan interval $(t_2 - t_1)$, i.e.,

$$change(I^{j,t_1}, I^{j,t_2}) \sim \Delta t_{12}, \text{ where } \Delta t_{12} = t_2 - t_1.$$

The main idea of DeepAtrophy (Dong et al., 2021) is to train a network to infer time-related measures from pairs of same-subject longitudinal images, which, under the assumptions above, would sensitize the network to progressive changes. However, directly inferring measures such as the interscan interval from a pair of scans is problematic because the rate of change in brain structures differs between individuals due to age, disease, inter-subject variability (Cash et al., 2015; Xie et al., 2020). A cognitively unimpaired older adult may experience less brain atrophy in two years as an individual with Alzheimer's disease in six months (Leow et al., 2009). Instead of predicting the interscan time interval in absolute terms, DeepAtrophy introduced two loss terms that incorporate interscan intervals in relative terms, are invariant to the individual rate of change, and are suitable to training deep learning networks. These two terms are the Scan Temporal Order (STO) loss and the Relative Interscan Interval (RISI) loss. The STO loss is a binary categorical loss that predicts the temporal order for a pair of same-subject scans. During training, pairs of scans are input in arbitrary temporal order, and the network's task is to correctly determine which of the input scans has an earlier scan date. The RISI loss is a multi-class categorical loss that predicts the relative inter-scan interval between two pairs of scans from the same subject. During training, the network is provided two pairs of scans from the same individual, and the network's task is to correctly assign the ratio of inter-scan intervals between the first and the second pair of scans into a set of discrete categories[1]. Our conclusion from studying these losses in DeepAtrophy is that minimizing the STO loss sensitizes the network to the presence of progressive changes between scan pairs; while minimizing the RISI loss sensitizes the network to the amount of progressive change (Dong et al., 2021).

In RDA, we continue with the idea of using STO and RISI prediction to sensitize the network to progressive changes. However, we apply these losses in a different way. The architecture of DeepAtrophy is that of an image classification network (i.e., 3D ResNet (S. Chen et al., 2019)) and the STO and RISI losses are applied to activation scores in the last layer. These activation scores lack interpretability, i.e., it is unknown what regions and characteristics of the input scans are driving the prediction. In contrast, the architecture of RDA is that of a pretrained registration network (VoxelMorph) paired with an attention network (U-Net). The STO and RISI losses are applied to features derived by applying the deformation field, output by the pretrained VoxelMorph network, to expansion/shrinkage ROIs proposed by the attention network. This is analogous to conventional DBM, where measures of atrophy are derived by applying the deformation field to anatomical ROIs, e.g., hippocampus segmentation. However, the attention mechanism, trained to minimize the STO and RISI losses, has the additional flexibility to emphasize image regions where the deformation field is reflective of neurodegeneration, and deemphasize or ignore image regions where the deformation field is dominated by other, non-progressive factors, such as MRI motion artifacts.

---

[1] Intuitively, using a regression-type loss for relative interscan interval would make more sense; however, in developing DeepAtrophy, we found such a formulation of the RISI loss to have poor convergence behavior.

The attention map is implemented as a 3D U-Net (Çiçek et al., 2016; Ronneberger et al., 2015) and is analogous to a 3D image segmentation network. It takes the same inputs as the pretrained VoxelMorph network (i.e., fixed and moving MTL ROIs) and outputs three activation maps corresponding to expansion, shrinkage, and background regions. We model shrinkage and expansion separately, since the MTL region contains structures that shrink with aging and disease progression (e.g., hippocampus, amygdala, cortex, white matter) as well as structures that increase in volume (e.g., ventricles). The background region corresponds to portions of the image where there are no changes or where the deformation field primarily reflects non-progressive changes, e.g., due to MRI artifacts. To be able to integrate the attention map with the deformation field for each of the three labels, the generated attention map is expected to be binary. This is achieved by applying a SoftMax layer with a large enough temperature coefficient ($\tau = 100$) to the last layer of the U-Net.

To generate features input to the STO and RISI losses, the deformation field $\Phi$ generated by the pretrained VoxelMorph network is applied to the attention ROIs. Let $S_l^{j,t_1,t_2}$ denote the output of the SoftMax for subject $j$ with timepoint $t_1$ used as the first input to RDA and timepoint $t_2$ used as the second input, and where $l$ denotes the label, with possible values 1 (shrinkage ROI), $-1$ (expansion ROI) and 0 (background). To keep the same directionality with ALOHA and VoxelMorph hippocampus atrophy measurement, the shrinkage region was taken as positive direction. Let $\Phi^{j,t_1,t_2}$ denote the corresponding deformation field output by VoxelMorph. Let $V_{l,1}^{j,t_1,t_2}$ and $V_{l,2}^{j,t_1,t_2}$ denote the volume of the ROI with label $l$ in the fixed image (timepoint $t_1$) and moving image (timepoint $t_2$), respectively. These volumes are estimated as follows:

$$V_{l,1}^{j,t_1,t_2} = \int S_l^{j,t_1,t_2}(\mathbf{x})d\mathbf{x}$$
$$V_{l,2}^{j,t_1,t_2} = \int \Phi^{j,t_1,t_2^{-1}}(S_l^{j,t_1,t_2}(\mathbf{x}))d\mathbf{x},$$
$$l \in \{1,-1\}$$

Note that for any given set of $j$, $t_1$, $t_2$, $V_{l,1}^{j,t_1,t_2} \neq V_{l,2}^{j,t_2,t_1}$ because both branches of the network will generate different outputs when the timepoints $t_1$ and $t_2$ are input in different order. This necessitates the use of the somewhat cumbersome notation above to represent ROI volumes in moving and fixed image space.

The absolute change in volume induced by the deformation $\Phi^{j,t_1,t_2}$ on the shrinkage/expansion ROI $l$ is defined as

$$A_l^{j,t_1,t_2} = l \cdot (V_{l,2}^{j,t_1,t_2} - V_{l,1}^{j,t_1,t_2})$$

Note that when scans are input in correct temporal order, i.e., $t_2 > t_1$, under the assumptions above, we expect $A_l^{j,t_1,t_2}$ to be negative (whether $l = -1$ or $l = +1$) and when the scans are input in inverse temporal order, we expect $A_l^{j,t_1,t_2}$ to be positive. Let $y_l^{t_1,t_2} = \text{sign}(t_2 - t_1)$ represent the ground-truth label of STO loss, the STO loss for one image pair is implemented as the cross-entropy loss:

$$l_{STO_1} = \sum_{l \in \{1,-1\}} y_l^{t_1,t_2} \log(\text{Sigmoid}(A_l^{j,t_1,t_2}))$$

As in DeepAtrophy, training the network with the RISI loss involves using two pairs of longitudinal scans from the same subject and putting both pairs through two copies of the RDA network with shared weights. The two pairs of images are selected such that the scan times of one of the image pairs falls strictly with the scan times of the other image pair i.e., scan times for the four input images $t_1$, $t_2$, $t_3$, and $t_4$ satisfy $[\min(t_1,t_2), \max(t_1,t_2)] \subseteq [\min(t_3,t_4), \max(t_3,t_4)]$, for scan pair $(I^{j,t_3}, I^{j,t_4})$ having a longer interscan interval or $[\min(t_3,t_4), \max(t_3,t_4)] \subseteq [\min(t_1,t_2), \max(t_1,t_2)]$, for scan pair $(I^{j,t_1}, I^{j,t_2})$ having a longer interscan interval. Furthermore, we allow one but not both endpoints of the interscan intervals to coincide (i.e., the two pairs of scans may be formed by three or four distinct scans). Ordering between $(I^{j,t_1}, I^{j,t_2})$ and $(I^{j,t_3}, I^{j,t_4})$ can be arbitrary. Applying the RDA network to each pair yields two sets of absolute atrophy measurements, $A_l^{j,t_1,t_2}$ and $A_l^{j,t_3,t_4}$. The STO loss is computed separately for the shrinkage and expansion regions, and for both image pairs, resulting in four different STO terms. The full STO loss is represented as below, and equal weights are applied to the four STO terms during RDA training.

$$l_{STO} = \sum_{l \in \{1,-1\}} y_l^{t_1,t_2} \log(\text{Sigmoid}(A_l^{j,t_1,t_2})) + \sum_{l \in \{1,-1\}} y_l^{t_3,t_4} \log(\text{Sigmoid}(A_l^{j,t_3,t_4}))$$

Under the hypothesis that the magnitude of progressive change in brain structures is proportional to time, we would expect

$$\frac{A_l^{j,t_1,t_2}}{A_l^{j,t_3,t_4}} = \frac{t_2 - t_1}{t_4 - t_3}$$

In principle, it should be possible to use a regression loss to train the network to generate outputs that satisfy this relationship. However, in practice, we found that a regression loss had poor convergence (in DeepAtrophy and, again, in training RDA). Rather, as common for order learning (Frank & Hall, 2001; L. Li & Lin, 2007) and like DeepAtrophy, this regression is reformulated into a multi-class classification problem. Instead of directly predicting an exact interscan interval ratio, the absolute value of ground-truth interscan interval ratio $\frac{|t_2-t_1|}{|t_4-t_3|}$ is predicted and categorized into four classes corresponding to ranges [0, 0.5], [0.5, 1], [1, 2], and [2, inf].

To formulate RISI as a categorical loss, we need to construct a differentiable mapping between absolute volume change values, $A_l^{j,t_1,t_2}$ and $A_l^{j,t_3,t_4}$, and a vector of probability values (one-hot vector) corresponding to the discrete categories. Geometrically, the values $A_l^{j,t_1,t_2}$ and $A_l^{j,t_3,t_4}$ can be represented as a point in $\mathbb{R}^2$, and the four discrete categories can be represented as partitions of $\mathbb{R}^2$ by lines passing through the origin. For example, if $\frac{|A_l^{j,t_1,t_2}|}{|A_l^{j,t_3,t_4}|}$ is in the range [0, 0.5] (first category) then $2A_l^{j,t_1,t_2} + A_l^{j,t_3,t_4} > 0$, and $2A_l^{j,t_1,t_2} - A_l^{j,t_3,t_4} < 0$ (upper cone), or $2A_l^{j,t_1,t_2} + A_l^{j,t_3,t_4} < 0$ and $2A_l^{j,t_1,t_2} - A_l^{j,t_3,t_4} > 0$ (lower cone). The boundaries of the corresponding region are shown as dashed red lines in Figure 2-(a). Assigning points within this region value 1 and points outside of this region value 0, we would obtain a map that maps the pair $A_l^{j,t_1,t_2}, A_l^{j,t_3,t_4}$ to a binary value for this category. However, to make the map differentiable, we can instead use the sigmoid function, i.e.,

$$P_{[0,0.5]}\left(A_l^{j,t_1,t_2}, A_l^{j,t_3,t_4}\right) = \max\left(\sigma_\alpha\left(2A_l^{j,t_1,t_2} + A_l^{j,t_3,t_4}\right) \times \sigma_\alpha\left(-2A_l^{j,t_1,t_2} + A_l^{j,t_3,t_4}\right),\right.$$
$$\left.\sigma_\alpha\left(-2A_l^{j,t_1,t_2} - A_l^{j,t_3,t_4}\right) \times \sigma_\alpha\left(2A_l^{j,t_1,t_2} - A_l^{j,t_3,t_4}\right)\right),$$

where $\sigma_\alpha(x) = 1/(1 + e^{-\alpha x})$ and $\alpha$ is a scale parameter. The resulting smooth map for category [0, 0.5] is plotted in Figure 2-(a); and the corresponding maps for the other three categories are plotted in Figure 2-(b-d). A more detailed formulation of the other three categories can be referred to in Supplementary S1. Note that to maximize the effectiveness of RISI loss, hyper-parameter $a$ in the Sigmoid function to control the softness of the decision boundary should be carefully chosen. After mapping $A_l^{j,t_1,t_2}$ and $A_l^{j,t_3,t_4}$ to a one-hot vector, the cross-entropy loss is applied.

In the inference stage, there is no need to compute the STO and RISI losses, and only a single image pair $(I^{j,t_1}, I^{j,t_2})$ is input to a single copy of the RDA network to generate a pooled attention area change $A_{\text{pooled}}^{j,t_1,t_2}$. $A_{\text{pooled}}^{j,t_1,t_2}$ is calculated as a subtraction of the volume of the shrinkage area and the expansion area in the corresponding attention maps:

$$A_{\text{pooled}}^{j,t_1,t_2} = A_{+1}^{j,t_1,t_2} + A_{-1}^{j,t_1,t_2}$$

As a longitudinal volume change measure, $A_{\text{pooled}}^{j,t_1,t_2}$ serves as a biomarker for neurodegeneration analysis in RDA model for a single image pair.

Figure 2. Functions used to map a pair of absolute volume change values, $A_l^{j,t_1,t_2}$ and $A_l^{j,t_3,t_4}$ corresponding to two image pairs $(I^{j,t_1}, I^{j,t_2})$ and $(I^{j,t_3}, I^{j,t_4})$ into a vector of probability values corresponding to four categories (one-hot vector) during the computation of the Relative Interscan

Interval (RISI) loss. Red dashed lines are exact boundaries corresponding to each category; the heat maps are the actual differentiable mapping functions obtained by using the sigmoid. See text for details.

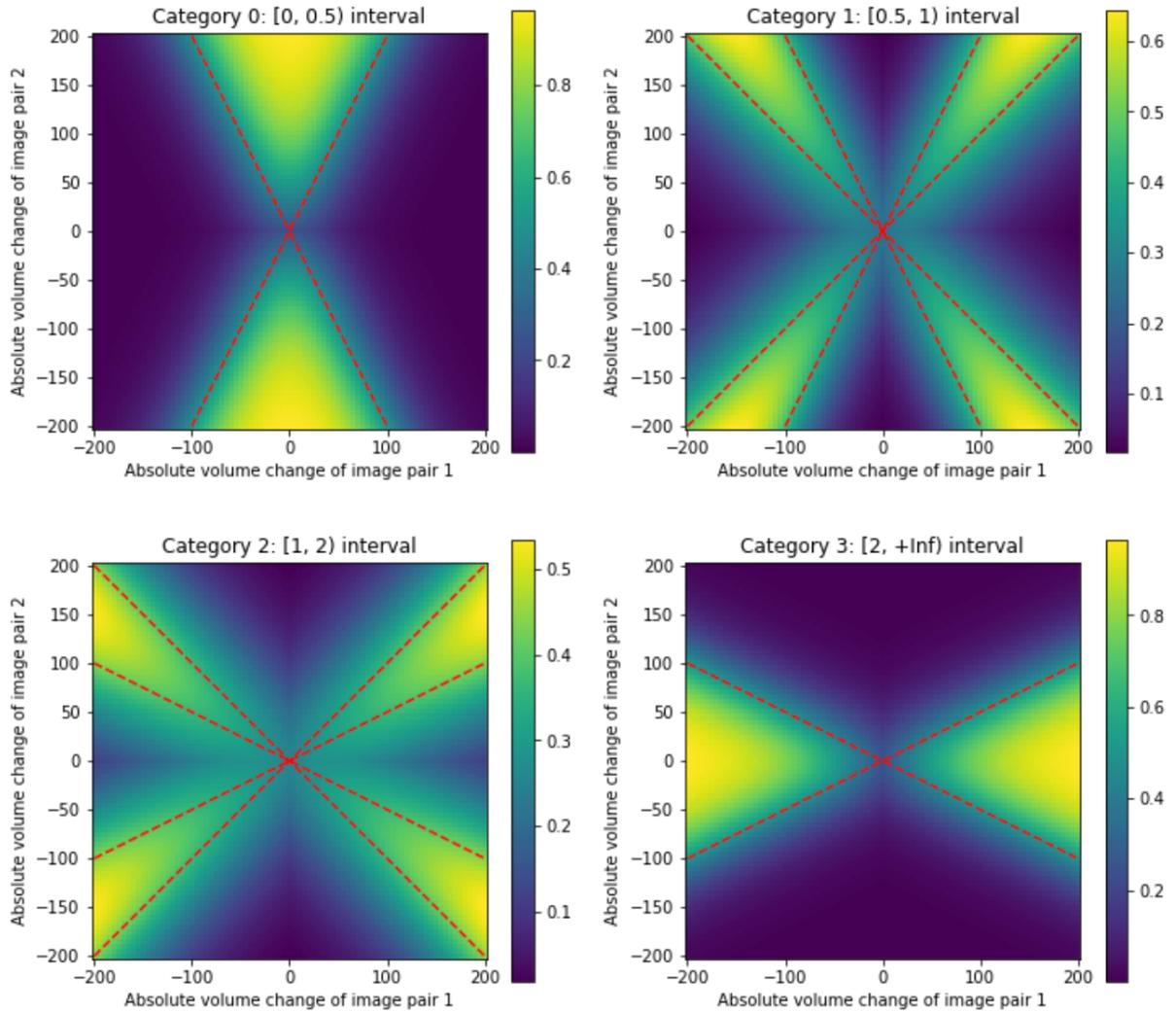

## Introduction of evaluation metrics

Five models are compared in Section 3. DBM method ALOHA (Das et al., 2012) and DeepAtrophy (Dong et al., 2021) were selected as baseline models. A DBM implementation based on VoxelMorph and RDA were compared to the baseline models to evaluate model performance. In addition, an ablation study was conducted where the attention map was predicted with only STO loss training but not RISI loss, namely $RDA_{STO\_only}$, to test the effectiveness of the RISI loss component in RDA. In all compared models, the same train/evaluation/test split and the same image pairs were selected in the test stage.

In four models, ALOHA, VoxelMorph, $RDA_{STO\_only}$, and RDA, the rate of volumetric change in regions of interest (referred to as atrophy rate, although its calculation differs in models, described as below) was computed as a measure of disease progression rate. In ALOHA and

VoxelMorph, atrophy rate was calculated as the volumetric difference of the whole hippocampus area (segmented by ASHS-T1) of a pair of longitudinal images divided by the first image (ordering is the same as the input to RDA model in the test stage). In RDA$_{STO\_only}$ and RDA, rather than using a pre-calculated hippocampus segmentation, a pooled attention area change $A_{pooled}^{j,t_1,t_2}$ was calculated as a subtraction of the volume of the shrinkage area and the expansion area in the corresponding attention maps. All other computations for atrophy rate were kept the same as in ALOHA and VoxelMorph computation. These four models all took shrinkage regions as positive direction. In DeepAtrophy, Predicted Interscan Interval (PII), as a measurement of predicted brain aging in comparison to actual interscan interval (assuming the same rate of disease and/or age progression for all individuals), was extracted as a measurement of the total amount of brain change (progression) in the MTL area. In addition, a single score Predicted-to-Actual Interscan Interval Rate (PAIIR) for each longitudinal image pair was derived as a surrogate biomarker of the rate of disease progression (Dong et al., 2021), and was compared with the other four methods. PAIIR is expected to be equal to 1 in the A- CU group, and larger with increasing clinical stages of AD.

Scan Temporal Order (STO) and Relative Inter-Sscan Interval (RISI) inference accuracy

The first set of evaluation metrics focuses on the ability of the five models to correctly infer temporal information from longitudinal scans that are in arbitrary temporal order. These evaluation metrics, denoted "STO accuracy" and "RISI accuracy", correspond to the STO and RISI losses used to train DeepAtrophy and RDA. They are evaluated on the test set, i.e., participants whose scans were unseen by the models during training. STO accuracy is measured as the fraction of scan pairs for which the sign of the change measurement output by the model (PII for DeepAtrophy, hippocampal volume change for ALOHA and VoxelMorph, pooled attention area change $A_{pooled}^{j,t_1,t_2}$ for RDA variants) matched the sign of the true interscan interval.

RISI accuracy is calculated as the fraction of experiments in which two pairs of scans from the same subject were input to a model, and the absolute value of the change measurement for the scan pair with a longer interscan interval was greater than the absolute value of the change measurement for the scan pair with a shorter interscan interval. Only subjects from the test set with at least three scans were selected, and the shorter interscan interval must fall within the longer one, as per the RISI training input conditions in Section 2.3.

Receiver operating characteristics (ROC) curve and the area under ROC (AUC) were reported for both STO accuracy and RISI accuracy using five methods, and DeLong's test (Delong et al., 2012) with the "pROC" R package (Robin et al., 2011) was used to determine the significance of difference in AUC between RDA and the other four methods.

Group differences in rates of disease progression

To assess the effectiveness of various measurements in detecting differences in rates of disease progression between individuals on different stages of the AD continuum, we compared the effect sizes for group comparisons between three A+ groups (A+ CU, A+ eMCI, A+ lMCI) and the A- CU group using PAIIR for DeepAtrophy and annualized ROI volume change rate

(atrophy rate) for other methods. To "simulate" different clinical trial durations, this group difference comparison was conducted separately by selecting for each subject either all follow-up scans in the 180-to-400-day window from the baseline scan; or in the 400-to-800-day window from the baseline scan. The test set was reduced to 250 subjects for the 400-day experiments and 226 subjects for the 800-day experiments, using the same subjects and scans for all methods.

When conducting this group difference analysis, a single measurement was derived for each subject and method, regardless of the number of scans in two simulated clinical trial scenarios. For subjects with only two scans, the annualized ROI volume change rate was calculated as the ratio of the volume difference divided by the time elapsed between scans for all models except DeepAtrophy. This value was further divided by the first scan's volume to account for different brain sizes and prevent misleading results from comparing absolute volume differences. For subjects with multiple scans, a single measurement was used to represent an overall progression rate for each method. For DeepAtrophy, a linear model was fit using PII values from all available scans, and the slope was taken as the summary PAIIR for the subject. For other methods, a linear model was fit to volumetric measurements using scan pairs, and the slope represents an annualized volume change with a unit of mL. This slope was divided by the average baseline volume to get an annualized volume change rate, which was then corrected for age based on a groupwise linear model. An unpaired one-sided Student's t-test was then conducted to compare disease progression measurements between the A+ groups (A+ CU, A+ eMCI, A+ lMCI) and normal group (A- CU).

### Attention heatmaps of Regional Deep Atrophy

The RDA model predicted an attention map for each longitudinal scan pair in the MTL to show areas of "shrinkage", "expansion", and the background. To summarize the attention map and evaluate RDA's explanation of regions of progression, an average heatmap was created from the shrinkage and expansion areas across all longitudinal pairs in the test set. To bring all attention maps to an average heatmap space, unbiased population templates of bilateral MTL provided by ASHS-T1 (Xie, Wisse, et al., 2019) were used as a reference, and all the fixed images in RDA in the test set were warped to the template space through deformable registration using the greedy package (Yushkevich et al., 2016). The attention maps of each scan were also warped to the template space using the corresponding deformation fields. In the template space, a heatmap for each voxel was calculated as the frequency of it being identified as shrinkage/expansion area divided by the total number of test examples. For instance, the attention heatmap for shrinkage area was calculated as:

$$Shrinking\ Attention\ Heatmap = \frac{\sum_{\{I_{test}\}} shrinking\ areas}{\sum_{\{I_{test}\}} tested\ image\ areas}$$

A higher value in the shrinkage attention heatmap indicates the tissue area is strongly linked to atrophy according to RDA's prediction. The peripheral areas in the template space may have fewer registered voxels compared to central areas, as the test images were cropped around the

MTL segmentation during the data preparation step, and the deformable registration to the template space may not cover the entire template area.

### Implementation details

The RDA model was trained on a single NVidia 2080Ti GPU computer using 6,154 image pairs for VoxelMorph and RDA$_{STO\_only}$ training, and 71,350 two-image pairs for RISI loss in RDA and DeepAtrophy training. The learning rate was set to $1 \times 10^{-5}$. Data augmentation applied to the deep learning pipeline included intensity normalization, random cropping centered on the hippocampus, random flipping, random rotation, and random image patch erasing. Each pair of longitudinal images underwent the same transformation (for example, rotation in the same angle, cropping at the same position) to ensure a voxel-wise match. Alignment was not applied to two image pairs from the same subjects during RISI training for RDA. For VoxelMorph, RDA$_{STO\_only}$, and RDA, intensity normalization, random cropping, random flipping, and random rotation was used. An additional data augmentation technique of random image patch erasing (replacing a random small 3D image patch with Gaussian noise) was applied to the RDA$_{STO\_only}$ model to avoid overfitting to the STO loss, but not applied to RDA (when optimizing on RISI loss) as it may affect the relative volume change measure between two image pairs. Details of training parameters, such as batch size, number of epochs, and the number of training images) can be referred to in Supplementary Section S2.

The hyper-parameters in our model were set without extensive tuning to show the feasibility of RDA. For VoxelMorph, hyper-parameters were set mostly to their default values in the original paper (Dalca et al., 2019). The regularization loss was set to normalized cross correlation loss (NCC) with a weight of 0.85, and the data similarity loss was set to L2 loss with a weight of 0.15, as suggested in Hoopes et al. (2021). In RDA, the SoftMax function with a temperature parameter of 100 was applied to the predicted probabilities of three regions to produce a 0 or 1 prediction for each voxel. Other hyper-parameters such as the weight between the STO and RISI loss (1 and 1) and the number of classification categories (4) were set the same as in DeepAtrophy.

## Results

### Scan Temporal Order (STO) inference accuracy

Table 2 reports the mean STO accuracy for all five evaluated models on the test set. Figure 3 further plots the receiver operating characteristic (ROC) curves for STO accuracy for the five methods and reports area under the ROC curve (AUC). Models that explicitly optimize STO accuracy (DeepAtrophy, RDA$_{STO\_only}$, and RDA) have significantly higher accuracy compared to DBM models (ALOHA and VoxelMorph). RDA$_{STO\_only}$ has the best performance in almost all disease stage groups, and overall. Comparing RDA and RDA$_{STO\_only}$, training only on the STO loss results in a higher STO accuracy. Compared to VoxelMorph, ALOHA has slightly higher STO accuracy, but the difference is not significant.

Table 2. Average accuracy for five models in inferring scan temporal order of same subject scan pairs input in arbitrary order (STO Accuracy). For all methods, lower accuracies in less impaired groups are expected because these groups undergo less neurodegeneration. Abbreviations: ALOHA = Automatic Longitudinal Hippocampal Atrophy software/package; RDA = Regional Deep Atrophy; RISI = Relative Interscan Interval; A + /A- = β-amyloid positive/negative; CU = cognitively unimpaired older adults; eMCI = early mild cognitive impairment; lMCI = late mild cognitive impair.

|  | A- CU | A+ CU | A+ eMCI | A+ lMCI | Average |
|---|---|---|---|---|---|
| ALOHA | 73.6% | 75.4% | 77.2% | 82.4% | 76.7% |
| VoxelMorph | 69.5% | 71.7% | 83.0% | 80.4% | 75.8% |
| DeepAtrophy | 87.2% | 87.7% | 90.1% | 91.2% | 88.9% |
| RDA$_{STO\_only}$ | **91.9%** | **95.0%** | **95.0%** | **94.7%** | **94.0%** |
| RDA | 91.3% | 90.5% | 94.6% | 93.7% | 92.5% |

Figure 3. Area under the receiver operating characteristic (ROC) curve (AUC) for the scan temporal order (STO) inference experiments for all five models. This figure uses the same data as in Table 2. Greater AUC for DeepAtrophy, RDA, and RDA$_{STO\_only}$ indicates greater accuracy in inferring the temporal order of scans. Abbreviations: ALOHA = Automatic Longitudinal Hippocampal Atrophy software/package; RDA = Regional Deep Atrophy; RISI = Relative Interscan Interval; A + /A- = β-amyloid positive/negative; CU = cognitively unimpaired older adults; eMCI = early mild cognitive impairment; lMCI = late mild cognitive impair.

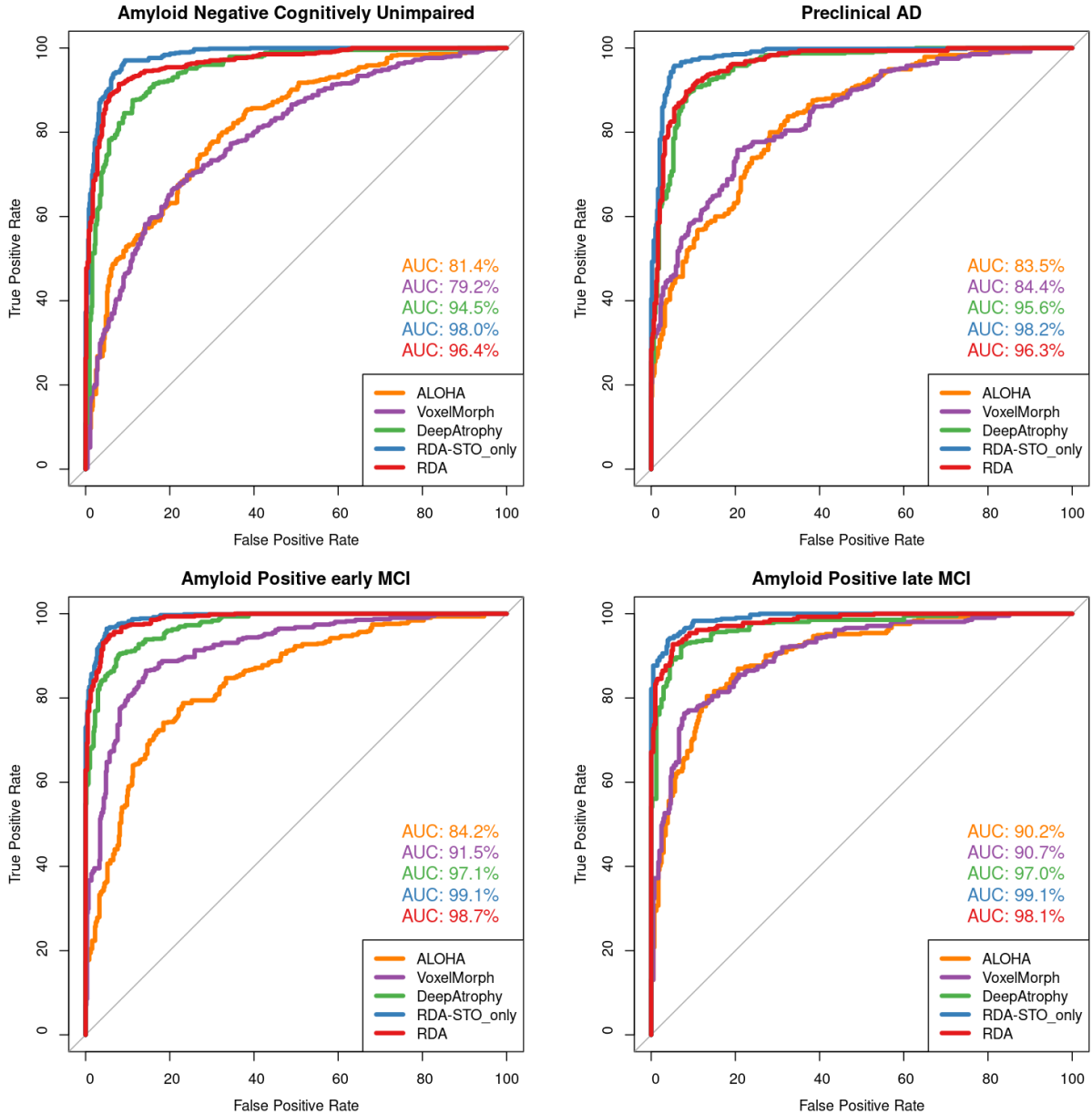

## Relative Inter-scan Interval (RISI) inference accuracy

    Table 3 compares the mean accuracy of five models in determining longer/shorter interscan interval (RISI accuracy) between two image pairs of the same subject. Models that explicitly optimize RISI accuracy (DeepAtrophy and RDA) have higher RISI accuracy compared to other models. RDA and DeepAtrophy achieve comparable RISI accuracy despite their different neural network architectures. Although only explicitly trained on STO loss, RDA$_{STO\_only}$ has improved RISI accuracy compared to VoxelMorph (81.0% vs. 71.6%), demonstrating complementary nature of STO and RISI loss in capturing time-related information in longitudinal volume change analysis.

    Figure 4 shows an AUC curve of RISI accuracy of different disease stages as depicted in Table 3. For all models, the accuracy of both RISI and STO (Table 2, 3) has an increasing trend as

disease severity go from A- CU (no disease) to A+ lMCI (most severe in our dataset). This outcome is predictable since as the disease worsens, the neurodegeneration is more pronounced in change detection compared to scanning noise.

Table 3. Comparison of Relative Inter-scan Interval (RISI) accuracy for five models. For two pairs of scans of from same subject, with the interscan interval of one scan pair strictly cover the other scan pair, we compare the accuracy of correctly identifying which scan pair has a longer interscan interval. For all methods, lower accuracies in less impaired groups are expected because less biological change happen in these subjects. Abbreviations: ALOHA = Automatic Longitudinal Hippocampal Atrophy software/package; RDA = Regional Deep Atrophy; RISI = Relative Interscan Interval; A + /A- = beta-amyloid positive/negative; CU = cognitively unimpaired adults; eMCI = early mild cognitive impairment; lMCI = late mild cognitive impair.

|  | A- CU | A+ CU | A+ eMCI | A+ lMCI | Average |
|---|---|---|---|---|---|
| ALOHA | 69.2% | 74.8% | 81.3% | 83.1% | 76.8% |
| VoxelMorph | 62.5% | 65.7% | 79.7% | 78.6% | 71.6% |
| Deep Atrophy | 81.1% | **88.7%** | 86.9% | **90.3%** | 86.1% |
| $RDA_{STO\_only}$ | 81.3% | 82.7% | 80.2% | 80.1% | 81.0% |
| RDA | **84.2%** | 88.2% | **89.0%** | 88.2% | **87.2%** |

Figure 4. Area under the receiver operating characteristic (ROC) curve (AUC) for relative interscan interval (RISI) accuracy experiments for all five models. This figure uses the same data as in Table 3. Greater AUC for DeepAtrophy, RDA, and $RDA_{STO\_only}$ indicates greater accuracy in inferring longer/shorter interscan interval for image pairs. Abbreviations: ALOHA = Automatic Longitudinal Hippocampal Atrophy software/package; RDA = Regional Deep Atrophy; RISI = Relative Interscan Interval; A + /A- = beta-amyloid positive/negative; CU = cognitively unimpaired adults; eMCI = early mild cognitive impairment; lMCI = late mild cognitive impair.

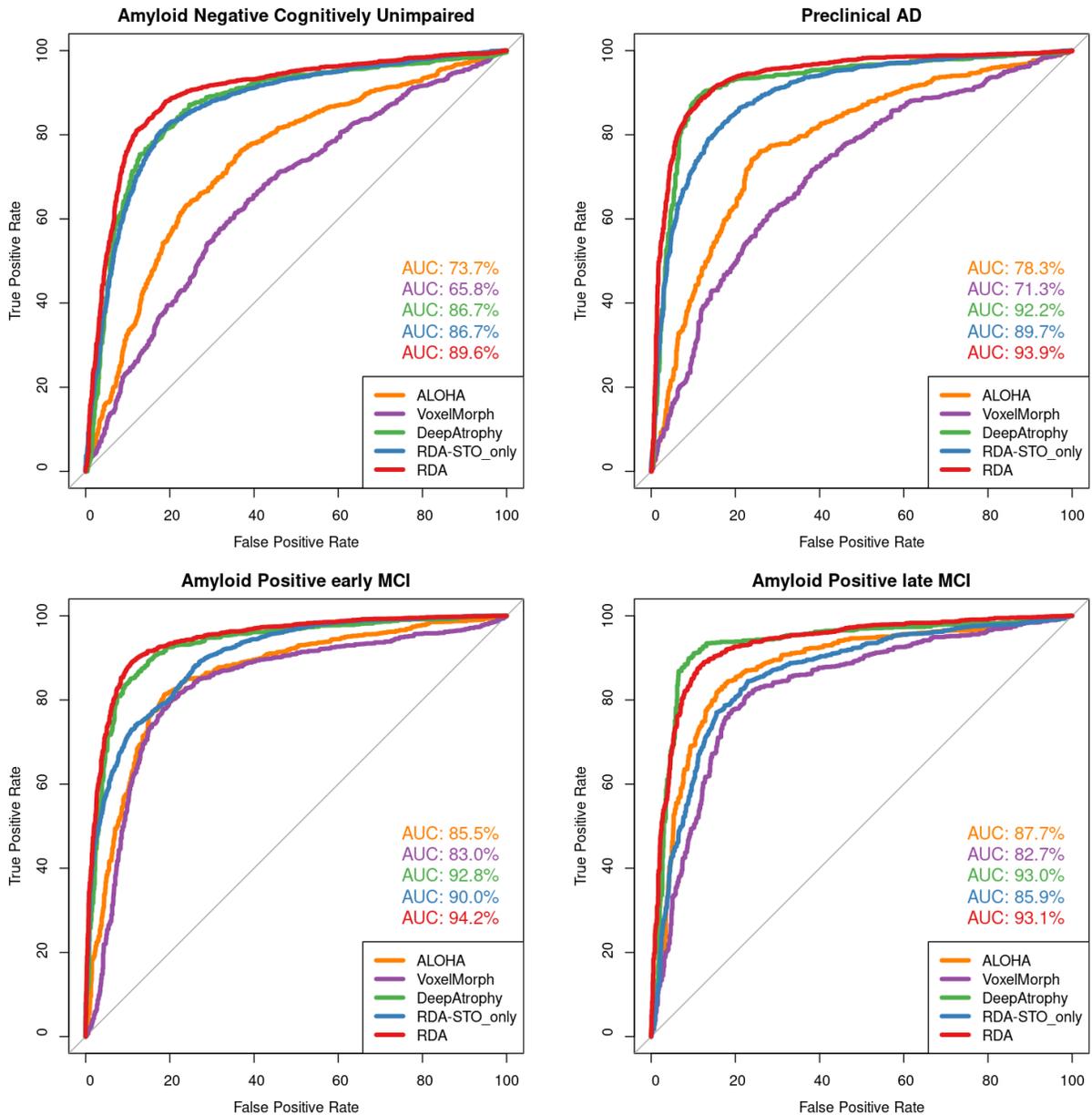

## Visualizing disease progression in individual subjects

In Figure 5, we plot the disease progression measurements of the five models for individual subjects over all scan times. For each subject and method, the change in measurement (PII for DeepAtrophy, hippocampal volume change for ALOHA & VoxelMorph, pooled attention area change for RDA & RDA$_{STO\_only}$) from the baseline scan to each follow up scan is displayed. The expected trend is to increase over time for PII, and to decrease over time for other methods (representing shrinkage of hippocampus in DBM or combined shrinkage in shrinkage areas and expansion in expansion areas in RDA variants). The relationship between brain structural progression and time is typically considered to be exponential (Jedynak et al., 2012). However, in practice some studies also simplify this model as a linear relationship (Cash et al., 2015; Das et al., 2012), since annual atrophy rates for normal aging elderly and Alzheimer's disease patients are relatively small (0.5% - 2% per year).

To evaluate the relationship between progression and time, we applied a non-parametric LOcal regrESSion (LOESS) model (Cleveland & Loader, 1996) to all the measurements (plotted in the bottom row of Figure 5), and expect a sub-linear relationship. Our results indicate that VoxelMorph and ALOHA have the most linear relationship, with higher sensitivity to time scale compared to other methods. For DeepAtrophy, $RDA_{STO\_only}$, and RDA, adding STO loss reduces false predictions (predicted expansion rather than shrinkage in ROIs), especially for scans with a longer interval. Adding RISI loss (comparing only RDA and $RDA_{STO\_only}$) results in a smoother/more linear relationship for RDA, especially for scans six-months apart (shown by the small red arrow). This suggests that adding the RISI loss enhances the network's understanding of disease progression in relationship to time.

Figure 5. Comparison of (a) DeepAtrophy Predicted Interscan Interval (PII); (b-e) VoxelMorph, $RDA_{STO\_only}$, RDA, and ALOHA volume change of different disease stages for individual subjects for all available scans. (f-j) are corresponding non-parametric fit (blue line) and confidence interval (grey area) of VoxelMorph, $RDA_{STO\_only}$, and RDA. For DeepAtrophy, PII is expected to be positive since the baseline scan. For all other methods, atrophy rate is expected to be negative since the baseline scan. Abbreviations: ALOHA = Automatic Longitudinal Hippocampal Atrophy software/package; RDA = Regional Deep Atrophy; RISI = Relative Interscan Interval; A+/A- = β-amyloid positive/negative; CU = cognitively unimpaired adults; eMCI = early mild cognitive impairment; lMCI = late mild cognitive impair.

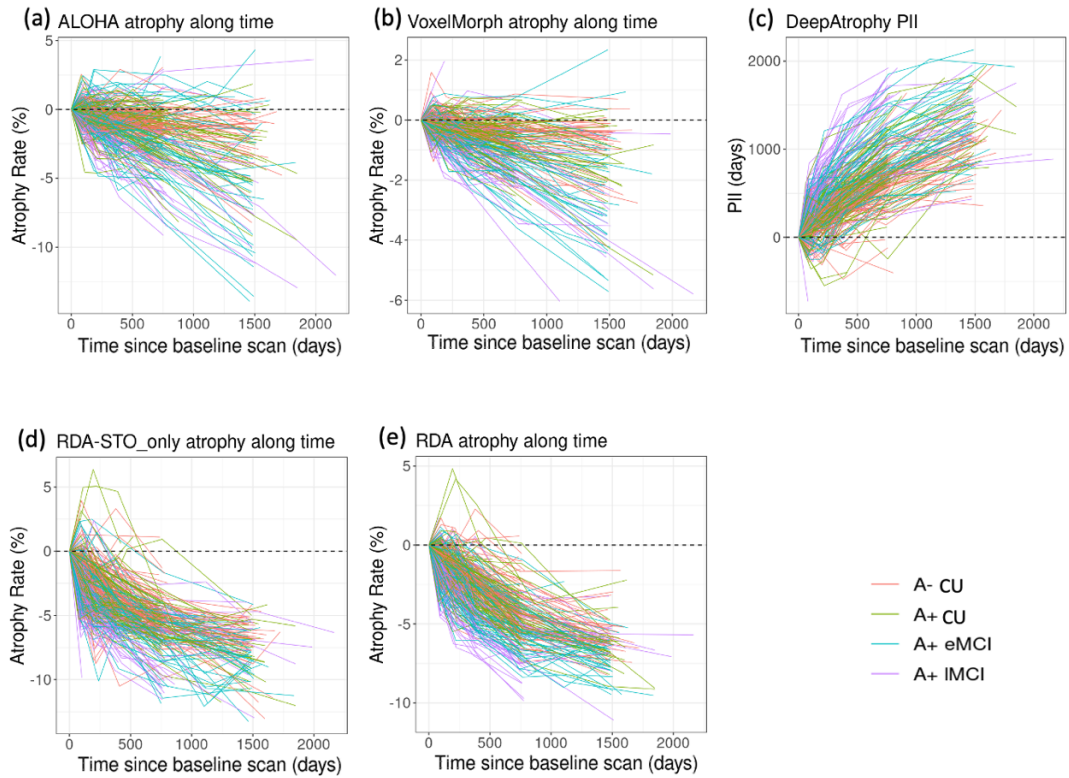

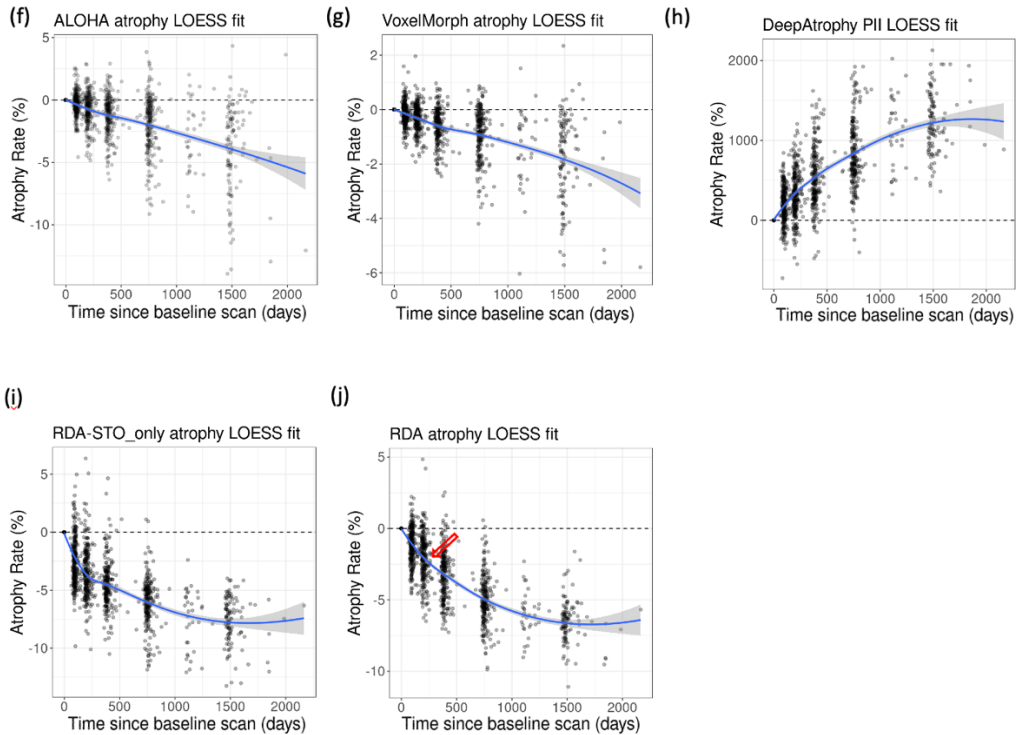

## Group differences in rates of disease progression

    In clinical trials aimed at slowing disease progression, it is crucial to evaluate the effectiveness of the treatment in the preclinical AD stage. Figure 6 demonstrates the ability of different measurements (PAIIR for DeepAtrophy and annualized atrophy rate for others) to detect group differences in progression over a follow-up period of 180 - 400 days (1 year) and 400 - 800 days (2 years). For all measurements, the sensitivity to disease stage is higher for scans with longer intervals (400-800 days) compared to shorter intervals (180-400 days). DeepAtrophy and RDA have a higher sensitivity in both the 180-400 days and 400-800 days analysis. Within 400 days, only DeepAtrophy and RDA can significantly distinguish between the normal control group (A- CU) and preclinical AD group (A+ CU) (p = 0.033 and p = 0.047 respectively, before False Discovery Rate correction), while other methods are not sensitive enough to detect differences with longitudinal MRI scans in such a short time. This highlights that RDA not only offers excellent interpretability, but also has the potential to detect subtle changes in a short amount of time between scans and to serve as a deep learning-based biomarker for longitudinal MRI measurements. Within 800 days, RDA does not have advantage over ALOHA in detecting difference in rates of progression at preclinical AD stage. This may be because RDA, even with training on RISI for atrophy rate measurement, is more sensitive to the direction of disease progression, rather than the magnitude, particularly for scans taken far apart (Figure 5, spaghetti plot). Without training on RISI loss (as in RDA$_{STO\_only}$), the effect of group differences is reduced and almost no difference can be detected between A- CU group and A+ CU group within 800 days.

Figure 6. Comparison of five models to detect differences in rates of progression from follow-up measurements (a) within 180 to 400 days and (b) within 400 to 800 days. For DeepAtrophy, age-adjusted Predicted-to-actual interscan interval rate (PAIIR), and for the rest four methods, age-adjusted annualized atrophy rate was applied to differentiate groups. For RDA and RDA$_{STO\_only}$, age-adjusted annualized atrophy rate was calculated from pooled attention area change (by subtracting shrinkage and expansion regions). In each subplot, the Wilcoxon signed-rank test was conducted to compare each patient group with the control group, and the p-values were shown for each comparison. Abbreviations: ALOHA = Automatic Longitudinal Hippocampal Atrophy software/package; RDA = Regional Deep Atrophy; RISI = Relative Interscan Interval; A+/A- = β-amyloid positive/negative; CU = cognitively unimpaired adults; eMCI = early mild cognitive impairment; lMCI = late mild cognitive impair.

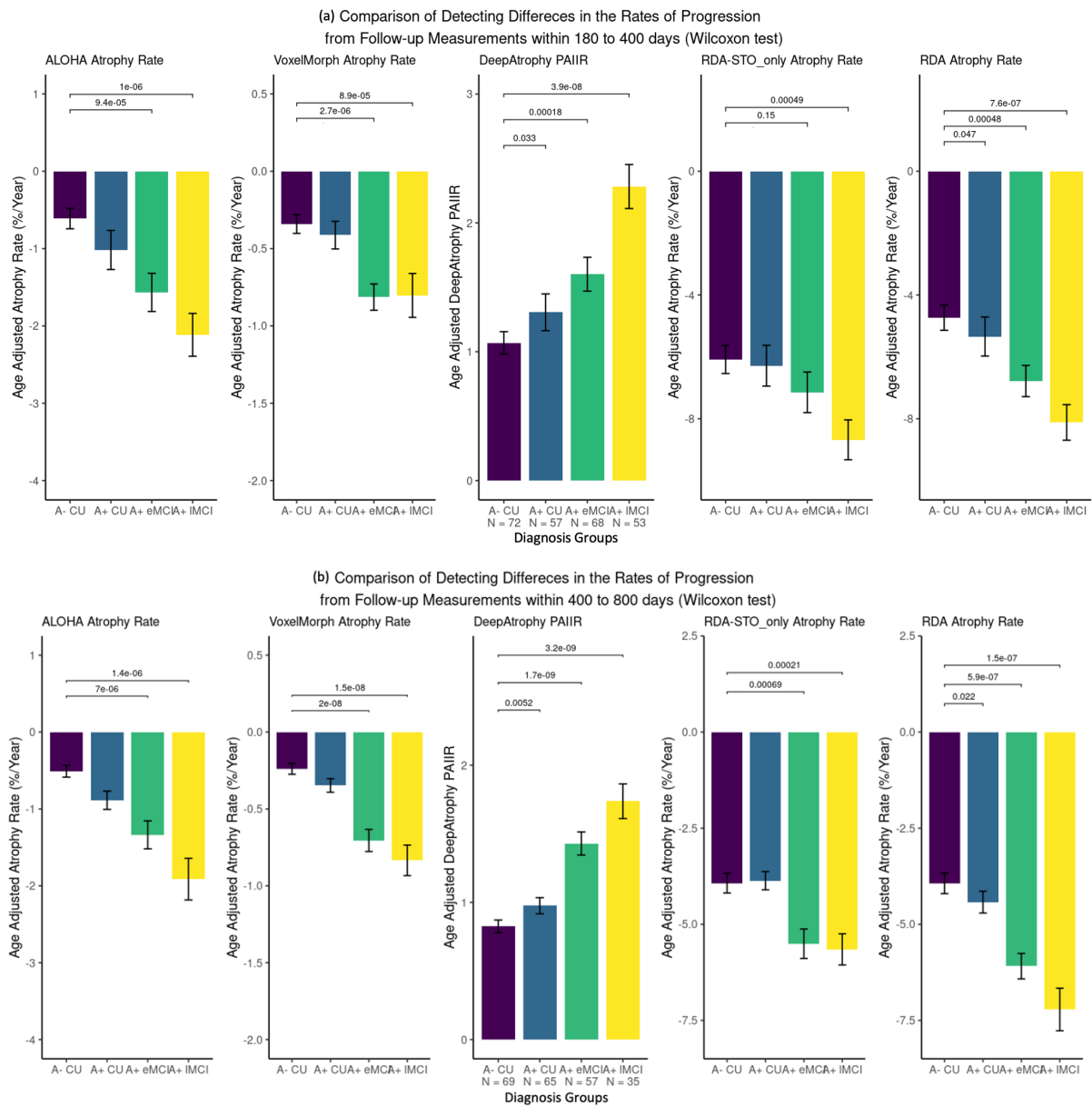

## Attention heatmaps of Regional Deep Atrophy

Figure 7 displays the shrinkage and expansion regions of RDA in (a-d) the MTL of example subjects and (e) a heatmap of averaged MTLs on the template space of all subjects. The heatmap shows that attention shrinkage regions are in the grey matter of the MTL and white matter of the parahippocampal gyrus, while expansion regions contain cerebrospinal fluid (CSF), such as ventricles. Despite lacking anatomical information in training, RDA's learned regions are roughly consistent with the anatomical structures associated with disease progression described in the literature (Braak & Braak, 1995; Platero et al., 2018). Unlike traditional disease progression analysis, which focuses on grey matter of the MTL or hippocampus, as this is the place that neuronal loss happens in the brain, RDA's predictions center more on white matter. This may be because (1) shrinkage regions in the image may be more heterogeneous, while a greater consensus can be observed in expansion regions, which are darker and have more uniform intensity (2) the deformation field generation in RDA was smoothed. In Figure 7 (a) where the attention map of a single subject was shown, shrinkage and expansion area are hardly adjacent. Background regions may indicate little or no structural change, or the biological change measurement is obscured by MRI noise and artifacts.

Figure 7. Regional Deep Atrophy (RDA) prediction of attention maps of (a)-(d) example subject scans after random flipping and (e) an average of all left MTLs in the test set. For a single subject (a)-(d), shrinkage area is overlayed on MRI images in green and expanding area is overlayed on MRI images in purple. For (e) an average of all MTLs, shrinkage area is represented in in hot colors and expanding area is represented in cold colors (see color-printed version for details).

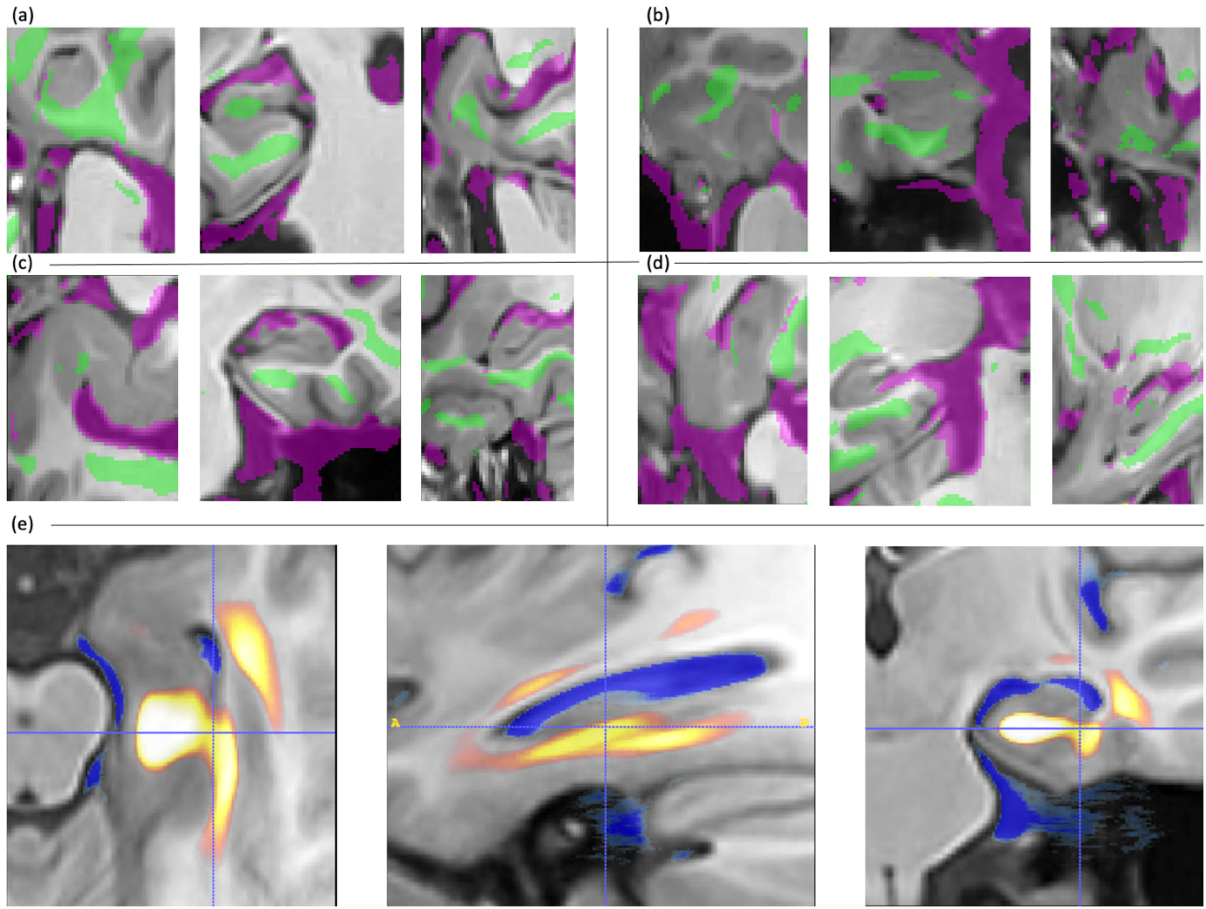

## Discussion

    In this study, we introduced a new deep learning model called Regional Deep Atrophy (RDA) that improves the interpretability of DeepAtrophy, for estimating neurodegeneration in longitudinal MRI scans. Our focus is on anatomical structures that are closely linked to early stages of AD progression, such as the MTL. Traditional methods (DBM/BSI) for estimating brain degeneration mainly apply image registration methods to estimate structural difference between scan pairs. Such analysis is sensitive to small shifts in structure boundaries, but suffers from limitations in accuracy due to imaging artifacts, such as subject motion, which may be misreported as neurodegeneration. DeepAtrophy reduces these limitations by applying losses that prime the network to detect time-related changes between image scans, hypothesizing that changes associated with the passing of time are due to neurodegeneration, while changes not associated with time are due to random factors, such as imaging artifacts. However, DeepAtrophy only provided a single summary measure of disease progression that lacks interpretability, making it difficult to determine if DeepAtrophy's prediction were due to MTL neurodegeneration or to other factors, such as changes in image texture, or expansion of the CSF space.

We thus designed a new deep learning framework RDA that combines the benefits of DeepAtrophy and DBM to increase the interpretability of DeepAtrophy's prediction, and associate this prediction with structural changes in the MTL area. The framework consists of the VoxelMorph model as a registration backbone, and a segmentation-like network (namely Attention Net) to identify regions sensitive to longitudinal changes. Adopting the same temporal information losses as DeepAtrophy, RDA reaches the same accuracy and group differentiation ability while also generating attention maps related to longitudinal structural change. Unlike DBM, which requires additional segmentation of ROIs for reporting atrophy rate measurements, RDA automatically detects regions in the scans that show longitudinal change, both shrinking and expansion, eliminating the need for external segmentation in atrophy measurement.

### Regional Deep Atrophy achieves high temporal inference accuracies

The results indicate that RDA and DeepAtrophy have similar temporal inference accuracies, around 90% and 85% respectively, for STO and RISI (Tables 2 and 3, Figures 3 and 4). On the other hand, registration-based methods VoxelMorph and ALOHA have lower accuracy of 71% to 77% for both STO and RISI. This implies that direct estimation of volumetric differences based on deformable registration in the hippocampal/MTL region, site of early changes in AD, may be inadequate due to imaging artifacts.

We relaxed ROIs from whole anatomical areas to automatically learned regions that are more sensitive to temporal information. Our hypothesis is that by training the model to detect temporal changes, it can focus on regions that are more related to time, and ignore random signals in the image that are not related to time, and thus provide sensitivity to progressive changes. Since some changes measured from registration methods may not be due to biological changes, but rather to noise, we infer that the learned attention map area is potentially more specific to biological changes in the brain. This hypothesis is supported by the observation that temporal inference accuracy generally increases with the severity of AD for all compared methods, indicating a positive correlation between biological changes in the brain and the ability of the models to detect these changes.

### Regional Deep Atrophy generated heatmaps correspond with anatomical structure

The attention maps generated by RDA are consistent with areas reported in previous literature to be sensitive to longitudinal change. AD related brain degeneration typically starts in the MTL, including the hippocampus and entorhinal cortex, and gradually spreads to cerebral cortex and the whole brain (Braak & Braak, 1995; Hua et al., 2016; Mueller et al., 2005; Vemuri & Jr, 2010). Hippocampus atrophy and ventricle expansion are commonly used as indicators of disease severity due to their close relationship with later cognitive functionality and higher consistency in tissue segmentation (Cash et al., 2015; Răzvan V. Marinescu et al., 2019). The average attention heatmap of RDA highlights shrinkage areas of the hippocampus, amygdala, MTL cortex and surrounding white matter, as well as an expanding area of CSF. These areas largely overlap with previously identified tissue change regions in the literature (Pettigrew et al., 2017; Wisse et al., 2014; Xie et al., 2020), although they may not strictly align with anatomical boundaries. Example attention maps also show that RDA detected shrinkage areas are farther from the boundary between CSF and grey matter, while expanding areas are tangent to it. This may be due to heterogeneity in tissue structures in MTL leading to smoothing in the deformation field and uncertainty when detecting shrinkage area in the attention map.

### VoxelMorph achieved comparable intra-subject registration accuracy to ALOHA

Our experiment on deep learning-based registration method VoxelMorph highlights its strong adaptability and accuracy when dealing with image pairs with subtle changes. It performs comparably or only slightly inferior to the DBM method ALOHA in terms of longitudinal MRI image registration. Compared to traditional image registration methods that optimize an objective function through solving differential equations, deep learning-based image registration methods can generate similarly accurate results in a faster manner, with deformation fields that retain important properties such as diffeomorphism. While we did not evaluate other deep learning registration methods (Fan et al., 2019b; Mok & Chung, 2021; Zhao et al., 2020, 2019), VoxelMorph remains one of the most widely used in practice.

### Regional Deep Atrophy generated biomarker maybe sensitive for early detection of AD

In clinical trials for early detection and intervention of AD, it is crucial to detect differences in biomarkers, such as atrophy rates between the placebo and treatment group, to evaluate treatment effectiveness, especially at the preclinical stage. We used bar plots to compare atrophy rates in different groups, with a focus on the preclinical AD group (A+ CU). Our results showed a near significance ($p < 0.05$) in RDA in differentiating preclinical AD group and the A- CU group, even when using data from a 180-400 day follow-up window, suggesting that these methods could also be effective biomarkers in the context of a drug trial focused on preclinical AD. As deep learning is a fast-emerging field and these are preliminary results, we believe both DeepAtrophy and RDA have the potential to lead to more sensitive disease progression biomarkers in future early AD clinical trials.

### Explainable deep learning for AD biomarker studies

While deep learning has been widely used in image-related fields, there is limited research on quantifying longitudinal atrophy in Alzheimer's disease using interpretable deep learning methods. Most previous studies (Hosseini-Asl et al., 2016; LULU YUE et al., 2019) have focused on classifying disease stages (e.g., AD vs. NC) through deep learning models. Of them, some models incorporate interpreting techniques, such as self-attention mechanisms (Jin et al., 2020), to the classification task. These models analyze whole brain MRI images (Jin et al., 2020; Lee et al., 2019; Qiu et al., 2022; Z. Tang et al., 2019) and generate heatmaps for each individual to highlight regions associated with AD. Most models show large areas and high probabilities (or p-values) in subjects with severe AD compared to cognitively normal individuals. Some models (Eitel et al., 2019; Zhang et al., 2021) ignore structural boundaries and learn a probability score for each voxel in the brain, while others (Zhu et al., 2022) use segmentations of different brain regions and learn the probability that a brain region is associated with the onset of AD. Our model focuses on the early onset of AD and learns to identify regions related to AD progression longitudinally. Our study is also unique in that it learns both shrinkage and expansion areas without overlap, that may both be associated with AD development. In the future, we can improve our analysis by using higher-resolution images and create a probability map for each subject instead of a nearly binary mask.

### Limitations and future work

RDA identifies anatomical regions around the hippocampus with accelerated structural change, including not only atrophy in white matter and grey matter, but also expansion in CSF and ventricles. One limitation of the model is that it only qualitatively shows heatmaps, but does not have a way to evaluate them quantitatively. It appears to focus more on white matter compared to grey matter regions. This may be because grey matter, as it shrinks with age, is anatomically closer to expanding CSF regions. This may cause difficulties for deep learning models in accurately drawing boundaries between grey matter and CSF, potentially missing some grey matter areas near the boundary. It may also be because the regularization of the deformation field in VoxelMorph registration causes the strongest signal to pool in central areas surrounded by grey matter, rather than individual cortical areas. Additional experiments focused on the regularization parameter in VoxelMorph may lead to more spatially focal attention maps.

Our model focuses on using deep learning techniques to automatically extract features, utilizing basic components such as U-Net and ResNet. With the rapid advancement in deep learning, more sophisticated modules, such as attention modules (Vaswani et al., 2017) and Vision Transformer (Dosovitskiy et al., 2020; Liu et al., 2021), have proved to be effective in various medical image analysis applications. In the future, we will investigate the possibility of incorporating these advanced architectures into our model. To train our model effectively, a large dataset is required. Indeed, we excluded preclinical AD from training to maximize the number of such individuals available for comparing trained models. With the increasing availability of longitudinal preclinical AD datasets, such as A4 dataset (Sperling et al., 2014), we plan to leverage these datasets to improve model performance and evaluate the efficacy of our model across different datasets.

## Conclusion

In this paper, we present Regional Deep Atrophy (RDA), a deep learning approach that can infer temporal information from same-subject longitudinal brain MRI scans. Like the earlier DeepAtrophy approach, the summary measures of change output by our method achieve excellent accuracy in ordering scans (newer/older) and interscan intervals (shorter/longer) compared to conventional registration-based methods; and can detect subtle differences in longitudinal change between groups at different stages of the AD continuum for both scans with one year and two years intervals, including between cognitively unimpaired individuals with and without PET-based evidence of AD pathology. Unlike DeepAtrophy, which only provided a single summary measure of change and lacked interpretability, RDA produces a dense deformation field that describes the longitudinal change across the imaged region and it also generates attention maps corresponding to regions where the deformation field captures progressive changes (atrophy or expansion) in the brain, as opposed to spurious changes, such as those caused by MRI artifacts. The atrophy/expansion regions identified by RDA roughly align with anatomical knowledge, though appear biased towards central white matter regions. The combination of image registration and deep learning-based temporal inference in RDA creates an interpretable and accurate deep learning biomarker for quantifying disease progression in AD clinical trials.

# Acknowledgment

This work was supported by National Institute of Health (NIH) (Grant Nos R01-AG056014, R01-AG040271, P30 AG072979, RF1-AG069474, R01-AG055005, R01-AG070592), and Foundation Philippe Chatrier. Data collection and sharing for this project was funded by the Alzheimer's Disease Neuroimaging Initiative (ADNI) (National Institutes of Health Grant U01 AG024904) and DOD ADNI (Department of Defense award number W81XWH-12–2–0012). ADNI is funded by the National Institute on Aging, the National Institute of Biomedical Imaging and Bioengineering, and through generous contributions from the following: AbbVie, Alzheimer's Association; Alzheimer's Drug Discovery Foundation; Araclon Biotech; BioClinica, Inc.; Biogen; Bristol-Myers Squibb Company; CereSpir, Inc.; Cogstate; Eisai Inc.; Elan Pharmaceuticals, Inc.; Eli Lilly and Company; EuroImmun; F. Hoffmann-La Roche Ltd and its affiliated company Genentech, Inc.; Fujirebio; GE Health- care; IXICO Ltd.; Janssen Alzheimer Immunotherapy Research & Development, LLC.; Johnson & Johnson Pharmaceutical Research & Development LLC.; Lumosity; Lundbeck; Merck & Co., Inc.; Meso Scale Diagnostics, LLC.; NeuroRx Research; Neurotrack Technologies; Novartis Phar- maceuticals Corporation; Pfizer Inc.; Piramal Imaging; Servier; Takeda Pharmaceutical Company; and Transition Therapeutics. The Canadian Institutes of Health Research is providing funds to support ADNI clinical sites in Canada. Private sector contributions are facilitated by the Foundation for the National Institutes of Health (www.fnih.org). The grantee organization is the Northern California Institute for Research and Education, and the study is coordinated by the Alzheimer's Therapeutic Research Institute at the University of Southern California. ADNI data are disseminated by the Laboratory for Neuro Imaging at the University of Southern California.

# Declaration of Interest

D.A.Wolk having received grants from Eli Lilly/Avid Radiopharmaceuticals, grants and personal fees from Merck, grants from Biogen, personal fees from Janssen, and personal fees from GE Healthcare; and L. Xie having received consulting fees from Galileo DSC, Inc. S.R. Das having received personal fees from Rancho Biosciences. None of the other authors has any financial relation to disclose.

Image Registration with Volume Tweening Network. *IEEE Journal of Biomedical and Health Informatics*, *24*(5), 1394–1404. https://doi.org/10.1109/JBHI.2019.2951024

Zhu, Y., Ma, J., Yuan, C., & Zhu, X. (2022). Interpretable learning based Dynamic Graph Convolutional Networks for Alzheimer's Disease analysis. *Information Fusion*, *77*, 53–61. https://doi.org/10.1016/j.inffus.2021.07.013